\documentclass{article}

\usepackage{amsmath, amssymb}
\usepackage{geometry}
\usepackage{soul}
\usepackage{textcomp}
\usepackage{paralist}
\usepackage{multirow}
\usepackage{subfigure}
\usepackage{graphics, graphicx}
\usepackage{epsfig}
\usepackage{float}
\usepackage[colorlinks,linkcolor=black,anchorcolor=black,citecolor=black]{hyperref}

\usepackage{threeparttable}

\newcommand\Tstrut{\rule{0pt}{2.6ex}}         
\newcommand\Bstrut{\rule[-1.3ex]{0pt}{0pt}}   

\newcommand\norm[1]{\lVert#1\rVert}

\providecommand{\keywords}[1]
{
	\small	
	\textbf{\textit{Keywords---}} #1
}

\title{Geometric Graph Learning with Extended Atom--Types Features for Protein--Ligand Binding Affinity Prediction}

\author{Md Masud Rana$^1$ and Duc Duy Nguyen$^1$ \footnote{Address correspondences to Duc Duy Nguyen. E-mail: ducnguyen@uky.edu}\\
$^1$ Department of Mathematics,
University of Kentucky, KY 40506, USA\\
}

\begin{document}

\maketitle

\begin{abstract}
Understanding and accurately predicting protein-ligand binding affinity are essential in the drug design and discovery process. At present, machine learning-based methodologies are gaining popularity as a means of predicting binding affinity due to their efficiency and accuracy, as well as the increasing availability of structural and binding affinity data for protein-ligand complexes. In biomolecular studies, graph theory has been widely applied since graphs can be used to model molecules or molecular complexes in a natural manner.
In the present work, we upgrade the graph-based learners for the study of protein-ligand interactions by integrating extensive atom types such as SYBYL and extended connectivity interactive features (ECIF) into multiscale weighted colored graphs (MWCG). By pairing with the gradient boosting decision tree (GBDT) machine learning algorithm, our approach results in two different methods, namely $^\text{sybyl}\text{GGL}$-Score and $^\text{ecif}\text{GGL}$-Score. Both of our models are extensively validated in their scoring power using three commonly used benchmark datasets in the drug design area, namely CASF-2007, CASF-2013, and CASF-2016. 
The performance of our best model $^\text{sybyl}\text{GGL}$-Score is compared with other state-of-the-art models in the binding affinity prediction for each benchmark. While both of our models achieve state-of-the-art results, the SYBYL atom-type model $^\text{sybyl}\text{GGL}$-Score outperforms other methods by a wide margin in all benchmarks.
\end{abstract}

\keywords{geometric graph learning, protein-ligand binding affinity, atom-type interaction, weighted colored subgraph, machine learning}


\section{Introduction}
In recent years, graph theories have been widely used in chemical, biological, physical, social, and computer sciences. This is because graphs are useful for representing and analyzing a wide range of practical problems. In molecular modeling, graph representation is widely used since it is a natural way to model their structures, in which graph vertices represent atoms and graph edges represent possible interactions between them. In general, graph theories can be divided into three categories: geometric graph theory, algebraic graph theory, and topological graph theory. Geometric graph theory studies a graph's geometric connectivity, which refers to the pairwise relations among graph nodes or vertices \cite{nguyen2017rigidity}. Algebraic graph theory concerns the algebraic connectivity via the characteristic polynomial, eigenvalues, and eigenvectors of matrices associated with the graph, such as the adjacency matrix or the Laplacian matrix \cite{nguyen2019agl, chen2021algebraic}. In topological graph theory, embedding and immersion of graphs are studied along with their association with topological spaces, such as abstract simplicial complexes \cite{wang2020persistent,meng2021persistent}.

There are numerous applications of graphs in chemical analysis and biomolecular modeling \cite{trinajstic2018chemical, schultz1989topological, janezic2015graph, angeleska2009dna}, such as normal-mode analysis (NMA) \cite{Go:1983, Tasumi:1982,brooks1983charmm,levitt1985protein} and elastic network model (ENM) \cite{bahar1997direct,flory1976statistical,bahar1998vibrational,atilgan2001anisotropy, hinsen1998analysis, tama2001conformational} used to study protein B--factor prediction.  Algebraic graph theory has been utilized in some of the most popular elastic network models (ENMs) such as the Gaussian network model (GNM) and the anisotropic network model (ANM). However, due to the matrix-diagonalization procedure, these methods have a computational complexity of $\mathcal{O}(N^3)$, with $N$ being the number of matrix elements. Furthermore, these methods suffer from limited accuracy in protein B--factor prediction, with average Pearson correlation coefficients less than 0.6 in all datasets. A geometric graph theory-based weighted graph approach, called flexibility-rigidity index (FRI), was introduced to bypass matrix diagonalization in GNM \cite{xia2013multiscale, opron2014fast, opron2016flexibility, nguyen2016generalized}. FRI assumes that protein interactions, including interactions with its environment, completely determine its structure in a given environment, which in turn, fully determines protein flexibility and functions. Therefore, it is not necessary to invoke a high-dimensional protein interaction Hamiltonian as in spectral graph theory to analyze protein flexibility when the accurate structure of the protein and its environment are known.  While the computational complexity of earlier FRI \cite{xia2013multiscale} is of $\mathcal{O}(N^2)$, the fast FRI \cite{opron2014fast} is of $\mathcal{O}(N)$. In order to capture multiscale interactions in macromolecules, multiscale FRI (mFRI) was introduced \cite{opron2015communication}, resulting in a number of graphs with parallel edges, i.e. multiple graphs.  Despite the fact that mFRI is about $20\%$ more accurate than the GNM on a set of 364 proteins, the average Pearson's correlation coefficient in B--factor prediction falls below 0.7, which is insufficient to provide a reliable assessment of protein flexibility. The limited accuracy of these graph-based models is due to the fact that they do not distinguish different chemical element types in a molecule or biomolecule, resulting in a severe loss of important chemical and biological information.

To address the aforementioned problem, a multiscale weighted colored graph (MWCG) model was introduced for protein flexibility analysis \cite{bramer2018multiscale}. In MWCG, the graph of a protein structure is colored according to the type of interaction between nodes in the graph, and subgraphs are defined according to colors. This process is commonly referred to as graph coloring, which is an important technique in graph theory that allows graph vertices or edges to be treated differently. 
MWCG weights the importance of graph edges by scaling their Euclidean distances in radial basis functions so that the nearest neighbors have the strongest edges in the sense of the Euclidean metric. Mathematical properties of MWCGs include low dimensionality, simplicity, robustness, and invariance of rotations, translations, and reflections.  Subgraphs constructed from vertex-labeled and edge-labeled graphs provide powerful representations of intermolecular and intramolecular interactions, such as hydrogen bonds, electrostatics, van der Waals interactions, hydrophilicity, hydrophobicity, etc \cite{nguyen2017rigidity, bramer2018multiscale}. The MWCG models offer 40\% more accuracy than the GNM in protein B--factor prediction \cite{bramer2018multiscale}.  

Molecular interactions between proteins and substrate molecules (ligands) are the principal determinant of many vital processes, such as cellular signaling, transcription, metabolism, and immunity. Therefore, understanding protein-ligand interactions is a central issue in biochemistry, biophysics, and molecular biology. Moreover, an accurate prediction of protein-ligand binding affinity plays a critical role in computational drug design, particularly in virtual screening and lead optimization. Various scoring functions (SFs) have been developed over the past few decades to evaluate protein-ligand interactions in structure-based drug design. These SFs can be classified mainly into four categories: force-field-based or physics-based SF, empirical SF, knowledge-based SF, and machine-learning-based SF. Force-field-based SFs offer physical insight and are not dependent on existing data. Empirical SFs utilizes a number of physical sub-models and use regression to fit existing data. 
The knowledge-based SF uses available datasets to derive binding patterns for proteins and ligands without requiring further training. 
Finally, machine learning-based SFs are data-driven, and are capable of capturing non-linear and complex relationships in the data. They can also easily handle large and diverse datasets. 
The performance of machine learning-based SFs strongly depends on the training set, in addition to their descriptors and machine learning algorithms. 
These scoring functions often take the top place in several standard benchmarks and community-wide competitions \cite{nguyen2019mathematical,nguyen2020mathdl,gaieb2018d3r,gaieb2018d3rGC3,parks2020d3r}. 

In recent years, due to the increasing availability of structural and binding affinity data for protein-ligand complexes, machine-learning SFs have become increasingly popular for binding affinity prediction. The RF--Score \cite{ballester2010machine} is considered one of the first machine-learning-based SFs to outperform other SFs in the CASF--2007 benchmark dataset. The model uses the random forest algorithm and employs element--type pair counts as features to describe protein-ligand complexes. The model was later extended to incorporate a more precise chemical description, including SYBYL atom-type pair counts features \cite{ballester2014does}. Including SYBYL atom types into the model permits deconvoluting the element into a hybridization state and bonding environment. For example, instead of having a single Carbon (C) element atom type, the SYBYL scheme allows the following subtypes: C.1, C.2 C.3, C.ar, and C.cat. A number of SYBYL atom-type-based models have been developed in the past years \cite{cheng2009comparative}, including SYBYL::ChemScore , SYBYL::G-Score , and SYBYL::D-Score. In a separate study, it has been shown that the connectivity of the atoms \cite{wojcikowski2019development} can improve the performance of a machine learning model in the binding affinity prediction task \cite{sanchez2021extended}. In \cite{sanchez2021extended}, the authors used a set of protein-ligand atom-type pair counts features, called the extended connectivity interactive features (ECIF), that considers each atom's connectivity to define the atoms involved in the pairs. The atom definition in ECIF is based on the atom environment concept initially introduced in the development of Extended Connectivity Fingerprints (ECFP) \cite{rogers2010extended}. Paired with a machine learning algorithm, the ECIF model significantly improves the performance of the binding affinity prediction with Pearson's correlation of 0.866 for the CASF--2016 benchmark. 
A number of machine-learning-based SF with different types of descriptors including differential geometry \cite{nguyen2019dg, rana2022eisa}, persistent homology \cite{cang2018integration, meng2021persistent}, and graph theory \cite{nguyen2017rigidity, nguyen2019agl} have emerged in the past few years for protein-ligand binding affinity prediction. Among them, the element-type graph coloring-based MWCG descriptors have particularly been successful in the task  \cite{nguyen2017rigidity, nguyen2019agl}. 

In the present work, we propose a geometric graph theory-based multiscale weighted colored graph (MWCG) descriptors for the protein-ligand complex where the graph coloring is based on SYBYL atom-type and ECIF atom-type connectivity. By pairing with the advanced machine learning architectures, our approach results in two different methods, namely $^\text{sybyl}\text{GGL}$-Score and $^\text{ecif}\text{GGL}$-Score. We verify the scoring power of our proposed model against three commonly used benchmarks in drug design, namely CASF-2007 \cite{cheng2009comparative}, CASF-2013 \cite{li2014comparative}, and CASF-2016 \cite{su2018comparative}. Several experiments confirm that both of our models achieve state-of-the-art results and outperform other models by a wide margin.






\section{Methods and Materials}\label{sec2}

\subsection{Multiscale Weighted Colored Geometric Subgraphs}

A graph $\mathcal{G}$ of a biomolecule consists of a set of vertices $\mathcal{V}$ and edges $\mathcal{E}$ and can be used to describe the noncovalent interaction of atoms in the molecule. In recent years, graph theory descriptors of protein-ligand binding interactions have been developed for massive and diverse datasets \cite{nguyen2019agl, jiang2021ggl}. To improve the graph theory representation, different types of elements are labeled, which is known as graph coloring. A colored graph is used to encode different types of interactions between atoms and gives rise to a basis for the collective coarse-grained description of the dataset. Labeled atoms of a molecule are classified into subgraphs where colored edges correspond to element-specific interactions.

To account for details of physical interactions in protein-ligand complexes such as hydrophobic, hydrophilic, etc., we are interested in constructing the subgraphs in an atomic interactive manner. In our previous work \cite{nguyen2017rigidity, nguyen2019agl}, we used the combination of the element symbols of the interacting protein-ligand atoms to classify the interaction,  e.g., C--C or N--O. In the present work, instead of the element symbol, we consider the following two schemes to classify the atomic interaction. 
In the first approach, we consider atom name (excluding hydrogen) for protein and SYBYL atom type for the ligand to define a range of protein-ligand atom pairs, e.g CD1--C.2, CG--C.ar, OE1--N.am, etc. In the second scheme, we consider the extended connectivity interaction features (ECIF) described in \cite{sanchez2021extended} to extract the protein-ligand atom-type pair that takes each atom's connectivity into account. The ECIF atom type in a molecule is defined by considering six atomic features: atom symbol, explicit valence, number of attached heavy atoms, number of attached hydrogens, aromaticity, and ring membership. Each of these properties can be represented textually where each property is separated by a semicolon. For example, the ECIF atom type for the $\alpha$ carbon CA  is C;4;3;1;0;0.

For convenience, let $\mathcal{T}$ be the set of all interested atom types in a given biomolecular dataset for either of the two schemes described above. To reduce the notation complexity, we denote the atom type at the $i$th position in the set $\mathcal{T}$ as $\mathcal{T}_i$. Assuming that a biomolecule has $N$ atoms of interest, we denote 
\begin{equation}
    \mathcal{V}=\{ (\mathbf{r}_i,\alpha_i)| \mathbf{r}_i\in \mathbb{R}^3; \alpha_i\in \mathcal{T}; i=1,2,\cdots,N\}
\end{equation} 
a subset of $N$ atoms (i.e. subgraph vertices) that are members of $\mathcal{T}$. Note that the $i$th atom is labeled by both its coordinate $\mathbf{r}_i$ and atom type $\alpha_i$. We assume that all the pairwise non-covalent interactions between atom types $\mathcal{T}_k$ and $\mathcal{T}_{k'}$ in a molecule or molecular complex can be represented by fast-decay weight functions 
\begin{align}
    \mathcal{E} &= \{ \Phi(\norm{\mathbf{r}_i-\mathbf{r}_j};\eta_{kk'})|  \alpha_i=\mathcal{T}_k,\, \alpha_j = \mathcal{T}_{k'}; \nonumber \\ 
    &\quad i,j=1,2,\cdots,N;\, \norm{\mathbf{r}_i-\mathbf{r}_j}\leq c \},
\end{align}
where $\lVert \mathbf{r}_i-\mathbf{r}_j \rVert$ is the Euclidean distance between the $i$th and $j$th atom and $c$ is a predefined cutoff distance that defines the binding site of the atom type $\mathcal{T}_k$ and $\mathcal{T}_{k'}$. Here $\eta_{kk'}$ is a characteristic distance between the atoms, and $\Phi$ is a subgraph weight that satisfies the following admissibility conditions

\begin{align}
    \Phi(\norm{\mathbf{r}_i-\mathbf{r}_j};\eta_{kk'}) &=1, \quad \mathrm{as}\; \norm{\mathbf{r}_i-\mathbf{r}_j}\rightarrow 0,\\
    \Phi(\norm{\mathbf{r}_i-\mathbf{r}_j};\eta_{kk'}) &=0, \quad \mathrm{as}\; \norm{\mathbf{r}_i-\mathbf{r}_j}\rightarrow \infty, \nonumber \\
    &\alpha_i=\mathcal{T}_k,\, \alpha_j = \mathcal{T}_{k'}.
\end{align} 

Although most radial basis functions can be used as the subgraph weight, the generalized exponential function 
\begin{equation*}
    \Phi_E(\norm{\mathbf{r}_i-\mathbf{r}_j};\eta_{kk'}) = e^{-(\norm{\mathbf{r}_i-\mathbf{r}_j}/\eta_{kk'})^\kappa}, \quad \kappa>0,
\end{equation*}
and the generalized Lorentz function
\begin{equation*}
    \Phi_L(\norm{\mathbf{r}_i-\mathbf{r}_j};\eta_{kk'}) = \frac{1}{1+\left(\norm{\mathbf{r}_i-\mathbf{r}_j}/\eta_{kk'}\right)^\kappa}, \quad \kappa>0,
\end{equation*}
 were shown to work very well for biomolecules \cite{opron2014fast}. Now, we have a weighted colored subgraph $G(\mathcal{V}, \mathcal{E})$ for a molecule or a molecular complex and we can use it to construct atomic-level collective molecular descriptors. We define the multiscale weighted colored geometric subgraph (MWCGS) interaction between $k$th atom type $\mathcal{T}_k$ and $k'$th atom type $\mathcal{T}_{k'}$ by
 \begin{align}\label{MWCGS_rigidity}
     \mu^G(\eta_{kk'}) &=\sum_i \mu_i^G(\eta_{kk'})=\sum_i \sum_j \Phi(\norm{\mathbf{r}_i-\mathbf{r}_j};\eta_{kk'}),\nonumber \\
     &\quad \alpha_i = \mathcal{T}_{k},\, \alpha_j = \mathcal{T}_{k'},
 \end{align}
where $\mu_i^G(\eta_{kk'})$ is the geometric subgraph centrality for the $i$th atom of type $\mathcal{T}_k$ and all atoms of type $\mathcal{T}_{k'}$. The summation over the geometric centrality $\mu_i^G(\eta_{kk'})$ in equation \eqref{MWCGS_rigidity} can be interpreted as the total interaction strength for the selected atom type pair $\mathcal{T}_k$ and $\mathcal{T}_{k'}$, which provides the atomic-level coarse-grained description of the molecular properties.
The equation \eqref{MWCGS_rigidity} is a generalization of a bipartite subgraph discussed in \cite{nguyen2017rigidity} for the predictions of protein-ligand binding affinities and free energy ranking. A bipartite subgraph of a protein-ligand complex is a graph in which each of its edges connects one atom in the protein and another in the ligand. We intend to capture the hydrogen bonds, polarization, electrostatics, van der Waals interactions, hydrophilicity, hydrophobicity, etc. of a protein-ligand complex through the bipartite graph coloring, i.e., atom-specific descriptions and subgraph weight.

\begin{figure}[htbp]
    \centering
  \includegraphics[width=0.7\textwidth]{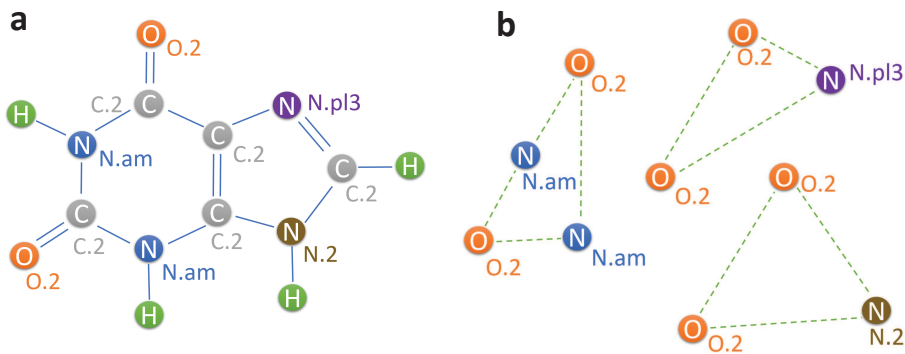}
    \caption{Illustration of the weighted colored subgraph. Part (a) is a diagram of the structure of the xanthine molecule ($\text{C}_5\text{H}_4\text{N}_4\text{O}_2$; ligand name: XAN; PDB ID: 2uz9), and (b) the weighted colored subgraphs, from left to right, $\text{G}_{\text{N.am--O.2}}$, $\text{G}_{\text{N.pl3--O.2}}$, and $\text{G}_{\text{N.2--O.2}}$  consisting of SYBYL atom-type pair N.am--O.2, N.pl3--O.2, and N.2--O.2, respectively. The dashed line in (b) represents the edges of the graph.}
    \label{fig:graph_cartoon}
\end{figure}

The multiscale behavior of the MWCGS arises when a different selection of the characteristic distance $\eta_{kk'}$ for a pair of atom types $k$ and $k'$ are considered. Therefore, for a molecule or a biomolecule, the MWCGS allows us to systematically construct a family of collective, scalable, multiscale graph-based descriptors by an appropriate selection of atom types pair $k$ and $k'$, characteristic distance $\eta_{kk'}$, and subgraph weight $\Phi$. An illustration of the weighted colored subgraph under the SYBYL atom-type system for the molecule xanthine ($\text{C}_5\text{H}_4\text{N}_4\text{O}_2$) is presented in Figure \ref{fig:graph_cartoon}.

\subsection{Geometric Graph Learning}
\begin{figure}[htbp]
    \centering
    \includegraphics[width=\textwidth]{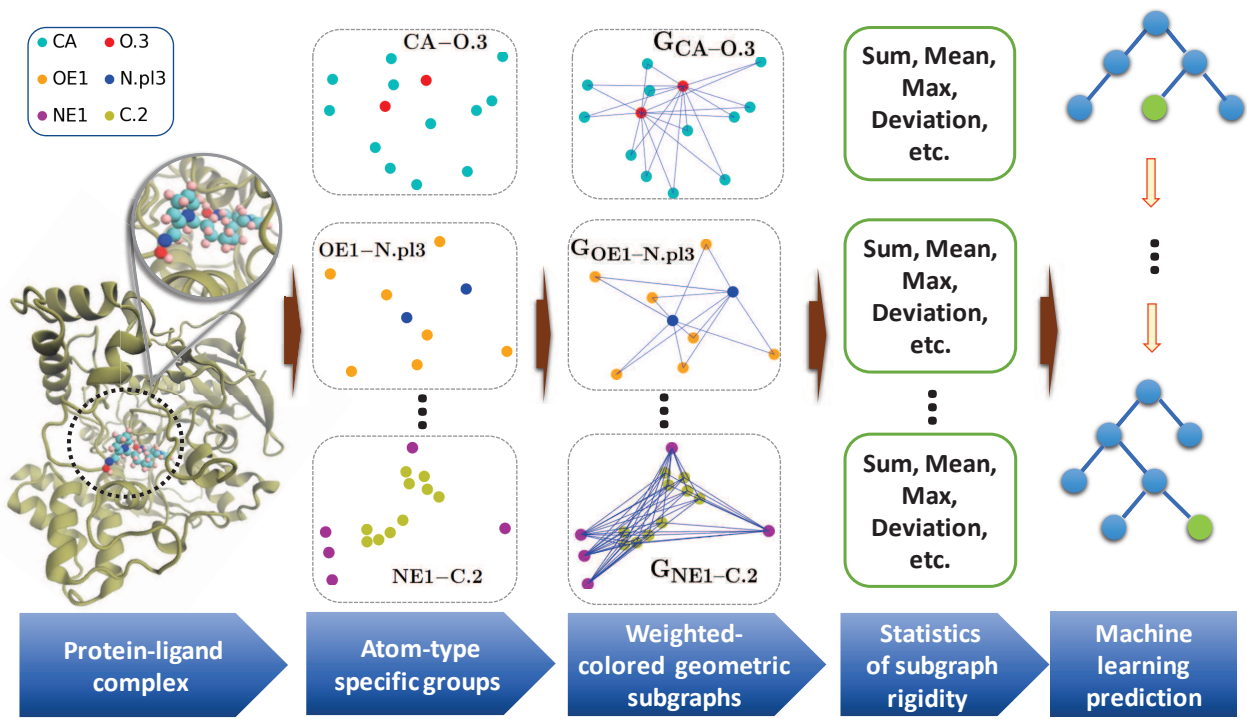}
    \caption{Illustration of the geometric graph learning strategy using the molecular complex with PDBID: 5bwc (first column). The second column represents the protein-ligand atom-type pair CA--O.3, OE1--N.pl3, and NE1--C.2, respectively from top to bottom. The corresponding weighted colored geometric subgraphs are shown in the third column. The fourth column presents the statistics of the subgraph rigidity. In the final column, the advanced machine learning models such as the gradient boosting trees integrate these statistical features for training and prediction.}
    \label{fig:ggl_strategy}
\end{figure}

The multiscale weighted colored geometric subgraph (MWCGS) descriptors for a molecule or molecular complex can be paired with any machine learning or deep learning algorithm to predict molecular properties. In a supervised machine learning algorithm (either classification or regression), the labeled dataset is divided into two parts: a training set and a test set. Let $\mathcal{X}_i$ be a labeled dataset for the $i$th molecule or molecular complex in the training set. Furthermore, suppose $\mathcal{G}(\mathcal{X}_i, \lambda)$ be a function that encodes the geometric information of the molecule or molecular complex into suitable graph representations with a set of parameters $\lambda$. The training of a machine learning model can be translated into a minimization problem,
\begin{equation}
    \min_{\lambda, \theta} \sum_{i\in I} \mathcal{L}(\mathbf{y}_i,\mathcal{G}(\mathcal{X}_i, \lambda); \theta)
\end{equation}
where $\mathcal{L}$ is a scalar loss function to be minimized and $\mathbf{y}_i$ is the labels of the $i$th sample in the training set $I$. Here, $\theta$ is the set of hyperparameters that depends on the chosen machine learning algorithm and typically be optimized for optimal performance. A wide range of machine learning algorithms, such as support vector machines, random forests, gradient boosting trees, artificial neural networks, and convolution neural networks, can be implemented in conjugation with the present geometric subgraph descriptors. However, to focus on the descriptive power of the proposed geometric subgraph features, we only employ gradient boosting decision trees (GBDT) in the present work and avoid optimizing machine learning algorithm selections. Although relatively simple, GBDT is still powerful, robust against overfitting,  and a widely used ensemble algorithm. 
An illustration of the proposed geometric graph learning strategy is presented in Figure \ref{fig:ggl_strategy}.

We use GBDT module in scikit-learn v0.24.1 package with the following parameters: $\texttt{n\_estimators}=20 000$, $\texttt{max\_depth} = 8$, \texttt{min\_samples\_split} = 2, \texttt{learning\_rate} = 0.005, \texttt{loss} = ls, \texttt{subsample} = 0.7, and \texttt{max\_features} = sqrt. These
parameter values are selected from the extensive tests on PDBbind datasets and are uniformly used in all our validation
tasks in this work. 

\subsection{Dataset}
For protein-ligand binding affinity prediction, the most commonly used benchmarks are the PDBbind datasets. In this work, we use the three most popular PDBbind benchmark datasets, CASF--2007, CASF--2013, and CASF--2016, to test the performance of our model. The PDBbind datasets consist of a general set, a refined set, and a core set, where the latter set is a subset of the previous one. In the present work, we explore two different training sets to build predictive models for the binding affinity of the complexes in the test set, which is the core set of the corresponding benchmark. The first training set, denoted by $S_R$, is the refined set excluding the core set of the corresponding benchmark. As a second training set, denoted by $S_G$, we use the general set excluding the core set of the corresponding benchmark. 
More information about these datasets is offered on the PDBbind website \url{http://pdbbind.org.cn/}. A summary of the dataset is provided in Table \ref{tab:PDBbind_dataset}.

\begin{table}[htbp]
\begin{center}
\caption{Summary of PDBbind datasets used in the present work.\Bstrut}
\begin{tabular}{llll}
	\hline
	Dataset & $\lvert S_G\rvert$ & $\lvert S_R \rvert$ & $\lvert S_C \rvert$\Tstrut\Bstrut\\
	\hline
    CASF--2007 benchmark & {2852} & 1105 & 195\Tstrut\Bstrut\\
    CASF--2013 benchmark & {11713} & 3516 & 195\Tstrut\Bstrut\\
    CASF--2016 benchmark & 12998 & 3772 & 285\Tstrut\Bstrut\\
    \hline
\end{tabular}
\label{tab:PDBbind_dataset}
\end{center}
\begin{tablenotes}
\item $\lvert S_G \rvert$: Number of complexes in the general set excluding the core set of the corresponding benchmark.
\item $\lvert S_R \rvert$: Number of complexes in the refined set excluding the core set of the corresponding benchmark.
\item $\lvert S_C \rvert$: Number of complexes in the core set of the corresponding benchmark.
\end{tablenotes}
\end{table}

\subsection{Model Parametrization}
For the sake of convenience, we use the notation $\text{GGL}_{\kappa, \tau}^{\alpha}$ to indicate the geometric graph learning features generated by using kernel type $\alpha$ and corresponding kernel parameters $\kappa$ and $\tau$. Here, $\alpha=E$ and $\alpha=L$ refer to the generalized exponential and generalized Lorentz kernels, respectively. 
And $\tau$ is used such that $\eta_{kk'}=\tau (\bar{r}_k + \bar{r}_{k'})$, where $\bar{r}_k$ and
$\bar{r}_{k'}$ are the van der Waals radii of atom type $k$ and atom type $k'$, respectively. Kernel parameters $\kappa$ and $\tau$ are selected
based on the cross-validation with a random split of the training data.
We propose a GGL representation in which multiple kernels are parametrized at different scale ($\eta$) values. In this work, we
consider at most two kernels. As a straightforward notation extension, two kernels can be parametrized by $\mathrm{GGL}_{\kappa_1,\tau_1;\kappa_2,\tau_2}^{\alpha_1,\alpha_2}$.
Each of these kernels gives rise to one set of features. Finally, as we consider two different schemes to extract the protein-ligand atom-type pair, we introduce the following two notations $^\text{sybyl}\text{GGL}_{\kappa_1,\tau_1;\kappa_2,\tau_2}^{\alpha_1,\alpha_2}$ and $^\text{ecif}\text{GGL}_{\kappa_1,\tau_1;\kappa_2,\tau_2}^{\alpha_1,\alpha_2}$.

\section{Results and Discussion}
In this section, we present the scoring power of our proposed geometric graph learning (GGL) model for the benchmark datasets discussed above. 


\subsection{Hyperparameter Optimization and Model Performance}
It is a well-known fact that the performance of a machine-learning model depends on the optimization of its essential parameters. To achieve the best performance of our GGL model on each benchmark, we optimize the kernel parameters $\kappa$ and $\tau$. We use five-fold cross-validation (CV) and a grid search method to find the optimal parameters $\tau$ in the range $[0.5,10]$ and $\kappa$ in the range $[0.5,20]$ with an increment of 0.5 for both parameter ranges. The high values of the power parameter $\kappa$ are considered to approximate the ideal low--pass filter (ILF) \cite{xia2015multiscale}. 

As a general strategy to optimize the model hyperparameters on each benchmark, we carry out a five-fold CV on the training set $S_R$ which is the refined set excluding the core set of the corresponding benchmark. Once we find the best model for each benchmark dataset, we test the performance of the model on the test set $S_C$ (i.e., the core set of the corresponding benchmark). For the prediction task, our first strategy is to train the model using the training set $S_R$ (i.e., the refined set excluding the core set) and observe the performance on the test set. And secondly, we train the best model using the training set $S_G$ (i.e., the general set excluding the core set) and test the performance on the test set. As the general set of each benchmark contains more diverse complexes than the refined set, we expect our model performs better when trained with the training set $S_G$. 
Below we discuss the optimization of our model hyperparameters $\tau$ and $\kappa$ and the model's performance on each benchmark. 

\subsubsection{CASF--2016}
\begin{figure}[htbp]
    \centering
    \includegraphics[width=1\linewidth]{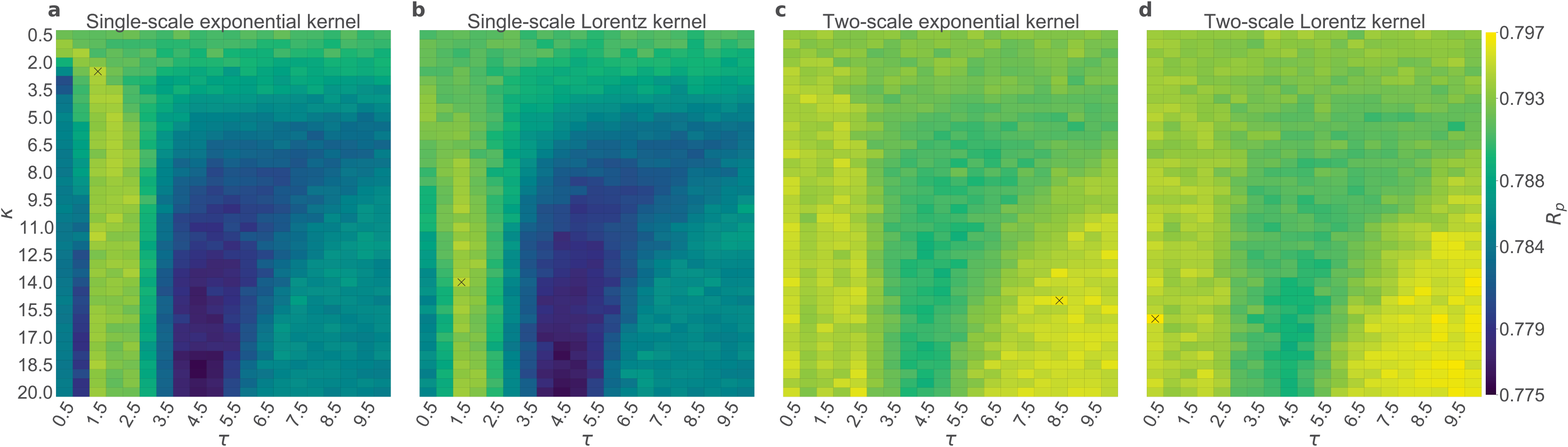}
    \caption{Optimized parameters for $^\text{sybyl}\text{GGL}$ model for CASF--2016 benchmark. The best parameters locations are marked by ``x''. The optimal parameters for, (a) single-scale exponential kernel model are $(\kappa,\tau)=(2.5,1.5)$ with the corresponding median $R_p=0.795$ and (b) single-scale Lorentz kernel model are $(\kappa,\tau)=(14.0,1.5)$ with corresponding median $R_p=0.795$. The optimal second kernel parameters for (c) two-scale exponential kernel model are $(\kappa,\tau)=(15.0,8.5)$ with the corresponding median $R_p=0.796$ and (d) two-scale Lorentz kernel model are $(\kappa,\tau)=(16.0,0.5)$ with the corresponding median $R_p=0.797$. }
    \label{fig:v2016_multitypes_kernel_params}
\end{figure}

The first benchmark we consider is the CASF--2016, the latest of the three benchmark datasets in the PDBbind database. We carry out five-fold cross-validation (CV) on the first training set which is the refined set excluding the core set of this benchmark. The CV results for both the single-scale and two-scale SYBYL atom-type GGL models are presented in Figure \ref{fig:v2016_multitypes_kernel_params}. The parameter set $(\kappa, \tau)=(2.5, 1.5)$ gives the best median Pearson's correlation coefficient $R_p$=0.795 for the single-scale exponential kernel (Figure \ref{fig:v2016_multitypes_kernel_params}a). For the single-scale Lorentz kernel model the parameters are $(\kappa, \tau)=(14.0, 1.5)$ with median $R_p$=0.795 (Figure \ref{fig:v2016_multitypes_kernel_params}b). The two-scale kernel model is built on top of the previously optimized single-scale kernel parameters, so we only optimize the parameters for the second kernel. Figure \ref{fig:v2016_multitypes_kernel_params}c and Figure \ref{fig:v2016_multitypes_kernel_params}d plots the CV results for the second kernel parameters $\kappa_2$ and $\tau_2$ of the two-scale kernel SYBYL atom-type model $^\text{sybyl}\text{GGL}_{\kappa_1,\tau_1;\kappa_2,\tau_2}^{\alpha_1,\alpha_2}$ with $\kappa_1$ and $\tau_1$ fixed at the optimal value from single-scale model. We observe that the best two-scale exponential kernel model is $^\text{sybyl}\text{GGL}_{2.5,1.5;15.0,8.5}^{\mathrm{E},\mathrm{E}}$ with median $R_p$=0.796 (Figure \ref{fig:v2016_multitypes_kernel_params}c) and the best two-scale Lorentz kernel model is $^\text{sybyl}\text{GGL}_{14.0,1.5;16.0,0.5}^{\mathrm{L},\mathrm{L}}$ with median $R_p$=0.797 (Figure \ref{fig:v2016_multitypes_kernel_params}d). 

\begin{figure}[htbp]
    \centering
    \includegraphics[width=1\linewidth]{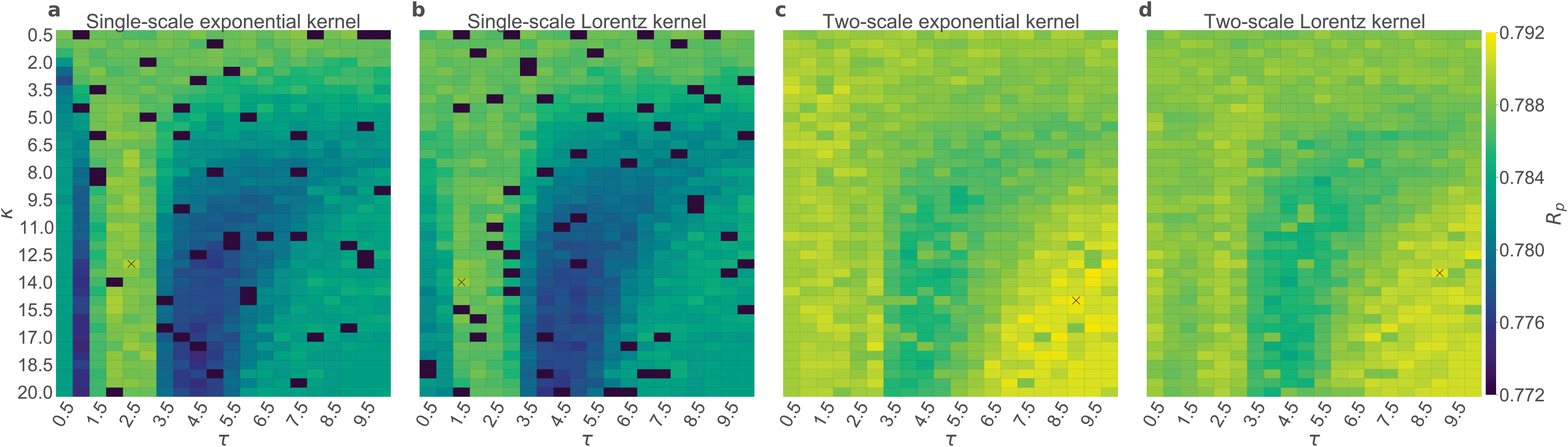}
    \caption{Optimized parameters for $^\text{ecif}\text{GGL}$ model for CASF--2016 benchmark. The best parameters locations are marked by ``x''. The optimal parameters for (a) single-scale exponential kernel model are $(\kappa,\tau)=(13.0,2.5)$ with the corresponding median $R_p=0.790$ and (b) single-scale Lorentz kernel model are $(\kappa,\tau)=(14.0,1.5)$ with corresponding median $R_p=0.789$. The optimal second kernel parameters for (c) two-scale exponential kernel model are $(\kappa,\tau)=(15.0,9.0)$ with the corresponding median $R_p=0.792$ and (d) two-scale Lorentz kernel model are $(\kappa,\tau)=(13.5,9.0)$ with the corresponding median $R_p=0.791$.}
    \label{fig:v2016_ecif_kernel_params}
\end{figure}

To find the optimal parameters for the ECIF atom-type models, we carry out a similar process discussed above. Figure \ref{fig:v2016_ecif_kernel_params} plots the CV performance of the single-scale kernel ECIF atom-type model $^\text{ecif}\text{GGL}_{\kappa,\tau}^{\alpha}$. We find that the best parameters for the single-scale exponential kernel model are $\kappa$=13.0 and $\tau$=2.5 with median $R_p$=0.790 (Figure \ref{fig:v2016_ecif_kernel_params}a) and the best parameters for the single-scale Lorentz kernel model are $\kappa$=14.0 and $\tau$=1.5 with median $R_p$=0.789 (Figure \ref{fig:v2016_ecif_kernel_params}b). The optimal parameters for the two-scale kernel model are also explored in a similar fashion as above. The CV results of each combination of the second kernel parameters are presented in Figure \ref{fig:v2016_ecif_kernel_params}. The figure confirms that the best two-scale exponential kernel model is $^\text{ecif}\text{GGL}_{13.0,2.5;15.0,9.0}^{\mathrm{E},\mathrm{E}}$ with median $R_p$=0.792 (Figure \ref{fig:v2016_ecif_kernel_params}c) and the best two-scale Lorentz kernel model is $^\text{ecif}\text{GGL}_{14.0,1.5;13.5,9.0}^{\mathrm{L},\mathrm{L}}$ with median $R_p$=0.791 (Figure \ref{fig:v2016_ecif_kernel_params}d). 

After finding the best models for this benchmark, we are interested in validating their performance on the test set, i.e., the CASF--2016 core set. The performance is measured using the Pearson\textquotesingle s correlation coefficient between the predicted and the experimental binding affinities of the test set complexes. First, we train each model with the smaller training set $S_R$, i.e., the PDBbind v2016 refined set excluding the CASF--2016 core set. Then we use the trained model to predict the test set. To this end, we repeat the model up to 50 times and use the average of all predicted values as the final predicted set. As a second approach, we train the model with the bigger training set $S_G$, i.e., the PDBbind v2016 general set excluding the CASF--2016 core set. For the prediction task, we again repeat the trained model 50 times and use the average of all predictions.   

The performance of the best models (both SYBYL atom-type and ECIF atom-type) on the test set are listed in Table \ref{tab:v2016_results}. We find that the performance of all models significantly improved when the model is trained with the bigger training data $S_G$. The results in Table \ref{tab:v2016_results} indicate that the two-scale models perform slightly better than their single-scale counterparts as expected. We also observe that the SYBYL atom-type models, both single-scale and two-scale, outperform their ECIF atom-type counterparts. The best model for this benchmark is the two-scale Lorentz kernel SYBYL atom-type model $^\text{sybyl}\text{GGL}_{14.0,1.5;16.0,0.5}^{\mathrm{L},\mathrm{L}}$ with reported Pearson\textquotesingle s correlation $R_p$=0.873. In addition, we compare the scoring power of our proposed GGL-Score against various state-of-the-art scoring functions in the literature \cite{cheng2009comparative,ballester2010machine,li2013id,li2015improving,li2014substituting,cao2014improved}. Figure \ref{fig:CASF_scoring_power}c illustrates such a comparison for CASF--2016 benchmark and clearly our model stands in the top. The second best is the TopBP-DL with reported $R_p$=0.848.  It must be stressed that the base geometric and algebraic graph learning models that consider the element-specific interactions instead of the atom-type interactions have a comparatively lower performance with $R_p$=0.815 \cite{nguyen2017rigidity} and $R_p$=0.835 \cite{nguyen2019agl} respectively. The above comparison and Figure \ref{fig:CASF_scoring_power}c confirm the scoring power and the effectiveness of considering atom-type pair interactions in the present model. 
Moreover, to highlight that the current model's impressive performance is due to the incorporation of the atom-type pair interactions  and not because of the use of larger training data $S_G$, we explore the performance of the base GGL models with element-specific interactions that are trained on the set $S_G$. The details of this experiment and results are presented in Appendix \ref{appendix}. While the use of the bigger training data improves the performance of the base GGL model, our extended atom-type models still outperform them by a big margin (see Table \ref{tab:standard_atom_type_results}).    


\begin{table}[htbp]
\begin{center}
\caption{Performance of various GGL models on CASF--2016 test set. \Bstrut}
\begin{tabular}{lcclc}
 \multicolumn{3}{@{}c@{}}{Pearson\textquotesingle s $R_p$ of single-scale Model} &
 \multicolumn{2}{@{}c@{}}{Pearson\textquotesingle s $R_p$ of two-scale Model}\Tstrut\Bstrut\\
 \cline{1-3}\cline{4-5}
   Model & Trained with $S_R$ & Trained with $S_G$ & Model & Trained with $S_G$ \Tstrut\Bstrut\\
	\hline
    $^\text{sybyl}\text{GGL}_{2.5,1.5}^{\mathrm{E}}$  & 0.838 &  0.872 & $^\text{sybyl}\text{GGL}_{2.5,1.5;15.0,8.5}^{\mathrm{E},\mathrm{E}}$ &  0.872\Tstrut\Bstrut\\
    $^\text{sybyl}\text{GGL}_{14.0,1.5}^{\mathrm{L}}$   & 0.832 &  0.872 & $^\text{sybyl}\text{GGL}_{14.0,1.5;16.0,0.5}^{\mathrm{L},\mathrm{L}}$ & 0.873\Tstrut\Bstrut\\  
    \hline
     $^\text{ecif}\text{GGL}_{13.0,2.5}^{\mathrm{E}}$  & 0.824 &  0.867 & $^\text{ecif}\text{GGL}_{13.0,2.5;15.0,9.0}^{\mathrm{E},\mathrm{E}}$ &  0.868\Tstrut\Bstrut\\
    $^\text{ecif}\text{GGL}_{14.0,1.5}^{\mathrm{L}}$   & 0.822 &  0.865 & $^\text{ecif}\text{GGL}_{14.0,1.5;13.5,9.0}^{\mathrm{L},\mathrm{L}}$ & 0.868\Tstrut\Bstrut\\
    \hline
\end{tabular}
\label{tab:v2016_results}
\end{center}
\end{table}

\subsubsection{CASF--2013}
\begin{figure}[htbp]
    \centering
    \includegraphics[width=1\linewidth]{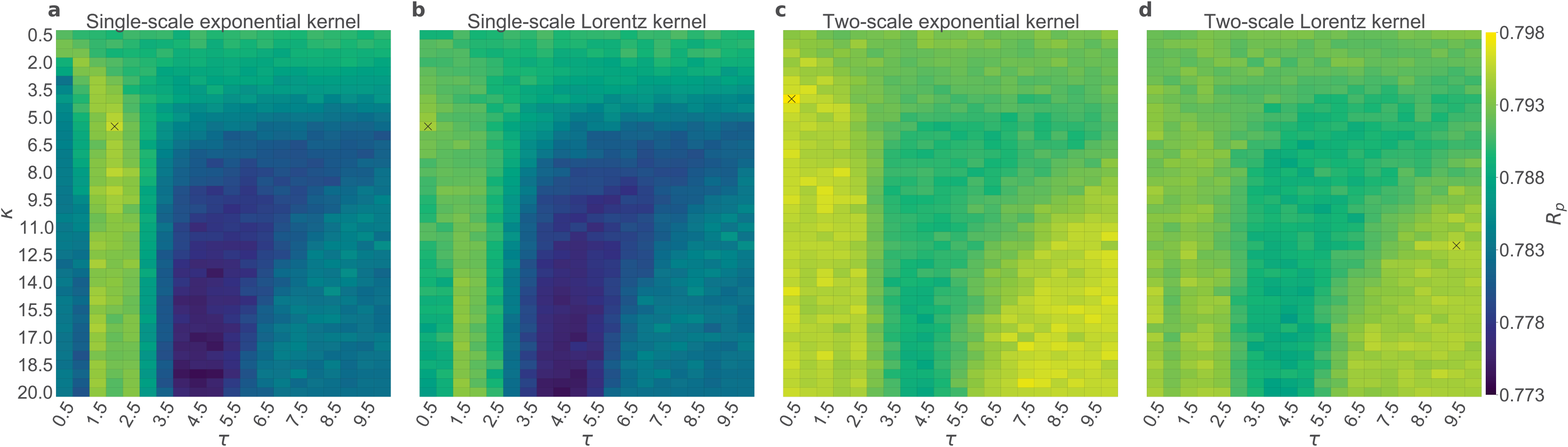}
    \caption{Optimized parameters for $^\text{sybyl}\text{GGL}$ model for CASF--2013 benchmark. The best parameters locations are marked by ``x''. The optimal parameters for (a) single-scale exponential kernel model are $(\kappa,\tau)=(5.5,2.0)$ with the corresponding median $R_p=0.796$ and (b) single-scale Lorentz kernel model are $(\kappa,\tau)=(5.5,0.5)$ with corresponding median $R_p=0.795$. The optimal second kernel parameters for (c) two-scale exponential kernel model are $(\kappa,\tau)=(4.0,0.5)$ with the corresponding median $R_p=0.798$ and (d) two-scale Lorentz kernel model are $(\kappa,\tau)=(12.0,9.5)$ with the corresponding median $R_p=0.798$.}
    \label{fig:v2013_multitypes_params}
\end{figure}

As a second benchmark dataset among the CASF family, we consider the CASF--2013 benchmark. For both SYBYL atom-type and ECIF atom-type models, we carry out a similar hyperparameter optimization to that of the CASF--2016 benchmark. We use the smaller training set $S_R$ of this benchmark which is the PDBbind v2015 refined set excluding the CASF--2013 core set for the cross-validation process. Figure \ref{fig:v2013_multitypes_params}  reveals the optimal parameters for the SYBYL atom-type model. The best parameters for the single-scale exponential kernel are found to be $\kappa$=5.5 and $\tau$=2.0 with median $R_p$=0.796 (Figure \ref{fig:v2013_multitypes_params}a) and the best parameters for the single-scale Lorentz kernel are $\kappa$=5.5 and $\tau$=0.5 with median $R_p$=0.795 (Figure \ref{fig:v2013_multitypes_params}b). For the two-scale kernel models, we  fix the first kernel parameters at their optimal value and optimize the second kernel parameter. Figure \ref{fig:v2013_multitypes_params}c shows that the best two-scale exponential kernel model is $^\text{sybyl}\text{GGL}_{5.5,2.0;4.0,0.5}^{\mathrm{E},\mathrm{E}}$ with median $R_p$=0.798 and Figure \ref{fig:v2013_multitypes_params}d shows that the best two-scale Lorentz kernel model is $^\text{sybyl}\text{GGL}_{5.5,0.5;12.0,9.5}^{\mathrm{L},\mathrm{L}}$ with median $R_p$=0.798. 

\begin{figure}[htbp]
    \centering
    \includegraphics[width=1\linewidth]{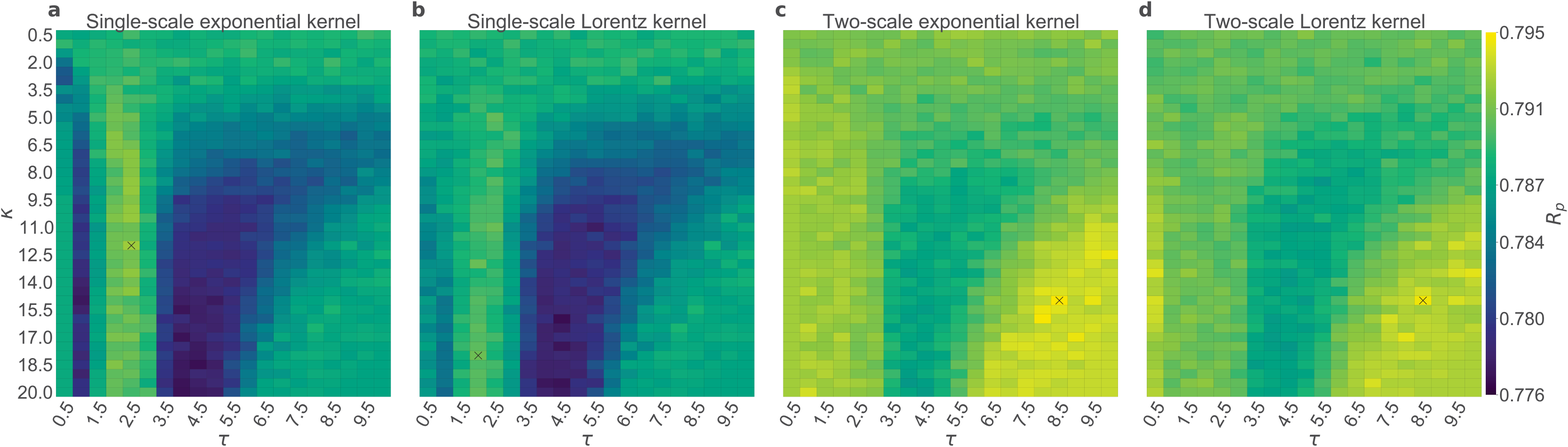}
    \caption{Optimized parameters for $^\text{ecif}\text{GGL}$ model for CASF--2013 benchmark. The best parameters locations are marked by ``x''. The optimal parameters for (a) single-scale exponential kernel model are $(\kappa,\tau)=(12.0,2.5)$ with the corresponding median $R_p=0.792$ and (b) single-scale Lorentz kernel model are $(\kappa,\tau)=(18.0,2.0)$ with corresponding median $R_p=0.791$. The optimal second kernel parameters for (c) two-scale exponential kernel model are $(\kappa,\tau)=(15.0,8.5)$ with the corresponding median $R_p=0.795$ and (d) two-scale Lorentz kernel model are $(\kappa,\tau)=(15.0,8.5)$ with the corresponding median $R_p=0.795$.}
    \label{fig:v2013_ecif_params}
\end{figure}

For the ECIF atom-type model hyperparameter optimization, we follow the same procedure as above. We found that the best single-scale exponential kernel model is $^\text{ecif}\text{GGL}_{12.0,2.5}^{\mathrm{E}}$ with median $R_p$=0.792 (Figure \ref{fig:v2013_ecif_params}a) and the best single-scale Lorentz kernel model is found to be $^\text{ecif}\text{GGL}_{18.0,2.0}^{\mathrm{L}}$ with median $R_p$=0.791 (Figure \ref{fig:v2013_ecif_params}b). For the two-scale kernel model, the best two-scale exponential kernel model is found to be  $^\text{ecif}\text{GGL}_{12.0,2.5;15.0,8.5}^{\mathrm{E},\mathrm{E}}$ with median $R_p$=0.795 (Figure \ref{fig:v2013_ecif_params}c). Finally, from Figure \ref{fig:v2013_ecif_params}d, we found that the best two-scale Lorentz kernel model is $^\text{ecif}\text{GGL}_{18.0,2.0;15.0,8.5}^{\mathrm{L},\mathrm{L}}$ with median $R_p$=0.795. 

Furthermore, we utilize the best models of this benchmark to predict the binding affinity of the 195 complexes in the CASF--2013 test set. Like the CASF--2016 benchmark, we first train each model using the smaller training set of this benchmark, i.e., the PDBbind v2015 refined set excluding the CASF--2013 core set, and then we generate a prediction for the test set from the average of 50 runs. Secondly, we use the more extensive training set, PDBbind v2015 general set, excluding the CASF--2013 core set to train the model and use it to get the prediction for the test set.

The performance of all models on the CASF--2013 test set is reported in Table \ref{tab:v2013_results}. It is interesting to see a similar trend that the performance of all models improved significantly when the model is trained on the bigger training data $S_G$. We also observe that the SYBYL atom-type models consistently outperform their ECIF atom-type counterparts. With the two-scale kernel model performing slightly better than the single-scale kernel model, the best-performing model for this benchmark is the two-scale exponential kernel SYBYL atom-type model $^\text{ecif}\text{GGL}_{12.0,2.5;15.0,8.5}^{\mathrm{E},\mathrm{E}}$ with reported Pearson\textquotesingle s correlation coefficient $R_p$=0.848. Additionally, Figure \ref{fig:CASF_scoring_power}b proves the dominance of our model in the scoring power over other published models for this benchmark. The reported $R_p$=0.848 of our best model is significantly higher than the $R_p$=0.808 of the runner-up model TopBP. Furthermore, Table \ref{tab:standard_atom_type_results} in Appendix \ref{appendix} confirms that the outstanding performance of our model is due to the incorporation of the atom-type interactions in the model. 

\begin{table}[htbp]
\begin{center}
\caption{Performance of various GGL models on CASF--2013 test set. \Bstrut}
\begin{tabular}{lcclc}
 \multicolumn{3}{@{}c@{}}{Pearson\textquotesingle s $R_p$ of single-scale Model} &
 \multicolumn{2}{@{}c@{}}{Pearson\textquotesingle s $R_p$ of two-scale Model}\Tstrut\Bstrut\\
 \cline{1-3}\cline{4-5}
   Model & Trained with $S_R$ & Trained with $S_G$ & Model & Trained with $S_G$ \Tstrut\Bstrut\\
	\hline
    $^\text{sybyl}\text{GGL}_{5.5,2.0}^{\mathrm{E}}$  & 0.797 &  0.846 & $^\text{sybyl}\text{GGL}_{5.5,2.0;4.0,0.5}^{\mathrm{E},\mathrm{E}}$ &  0.848\Tstrut\Bstrut\\
    $^\text{sybyl}\text{GGL}_{5.5,0.5}^{\mathrm{L}}$   & 0.812 &  0.841 & $^\text{sybyl}\text{GGL}_{5.5,0.5;12.0,9.5}^{\mathrm{L},\mathrm{L}}$ & 0.844\Tstrut\Bstrut\\  
    \hline
     $^\text{ecif}\text{GGL}_{12.0,2.5}^{\mathrm{E}}$  & 0.797 &  0.826 & $^\text{ecif}\text{GGL}_{12.0,2.5;15.0,8.5}^{\mathrm{E},\mathrm{E}}$ &  0.829\Tstrut\Bstrut\\
    $^\text{ecif}\text{GGL}_{18.0,2.0}^{\mathrm{L}}$   & 0.801 &  0.829 & $^\text{ecif}\text{GGL}_{18.0,2.0;15.0,8.5}^{\mathrm{L},\mathrm{L}}$ & 0.833\Tstrut\Bstrut\\
    \hline
\end{tabular}
\label{tab:v2013_results}
\end{center}
\end{table}

\subsubsection{CASF--2007}
\begin{figure}[htbp]
    \centering
    \includegraphics[width=1\linewidth]{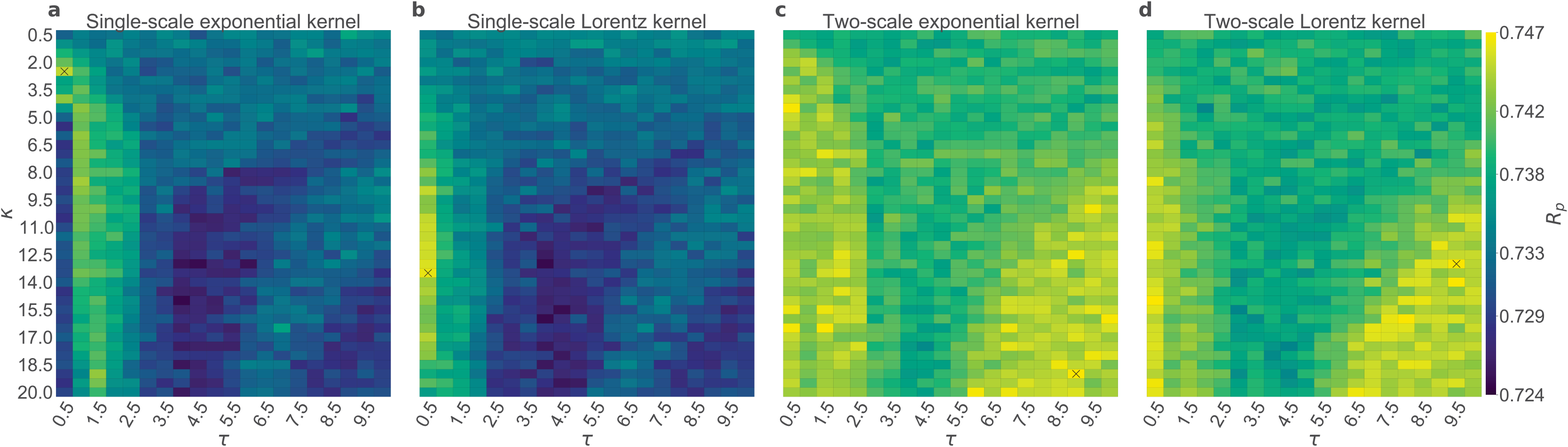}
    \caption{Optimized parameters for $^\text{sybyl}\text{GGL}$ model for CASF--2007 benchmark. The best parameters locations are marked by ``x''. The optimal parameters for (a) single-scale exponential kernel model are $(\kappa,\tau)=(2.5,0.5)$ with the corresponding median $R_p=0.745$ and (b) single-scale Lorentz kernel model are $(\kappa,\tau)=(13.5,0.5)$ with corresponding median $R_p=0.746$. The optimal second kernel parameters for (c) two-scale exponential kernel model are $(\kappa,\tau)=(19.0,9.0)$ with the corresponding median $R_p=0.747$ and (d) two-scale Lorentz kernel model are $(\kappa,\tau)=(13.0,9.5)$ with the corresponding median $R_p=0.747$.}
    \label{fig:v2007_multitypes_params}
\end{figure}

Our last benchmark is the CASF--2007. The hyperparameter optimization for this benchmark is similar to the previous two benchmarks. The smaller training set $S_R$, which is the PDBbind v2007 refined set excluding the CASF-2007 core set, is used for the five-fold CV. The CV performances of the SYBYL atom-type models are plotted in Figure \ref{fig:v2007_multitypes_params}.  The optimal kernel parameters for the single-scale exponential model are $\kappa$=2.5 and $\tau$=0.5 (Figure \ref{fig:v2007_multitypes_params}a) with median $R_p$=0.745. For the single-scale Lorentz kernel, Figure \ref{fig:v2007_multitypes_params}b, the best parameters are $\kappa$=13.5 and $\tau$=0.5 with median $R_p$=0.746. The two-scale models are built on top of the optimized single-scale model, we only search for the optimal second kernel parameters. Figure \ref{fig:v2007_multitypes_params}c shows that the two-scale exponential model  $^\text{sybyl}\text{GGL}_{2.5,0.5;19.0,9.0}^{\mathrm{E},\mathrm{E}}$ gives the best median $R_p$=0.747 while Figure \ref{fig:v2007_multitypes_params}d reveals that the best two-scale Lorentz kernel model is $^\text{sybyl}\text{GGL}_{13.5,0.5;13.0,9.5}^{\mathrm{L},\mathrm{L}}$ with median $R_p$ being 0.747. 

\begin{figure}[htbp]
    \centering
    \includegraphics[width=1\linewidth]{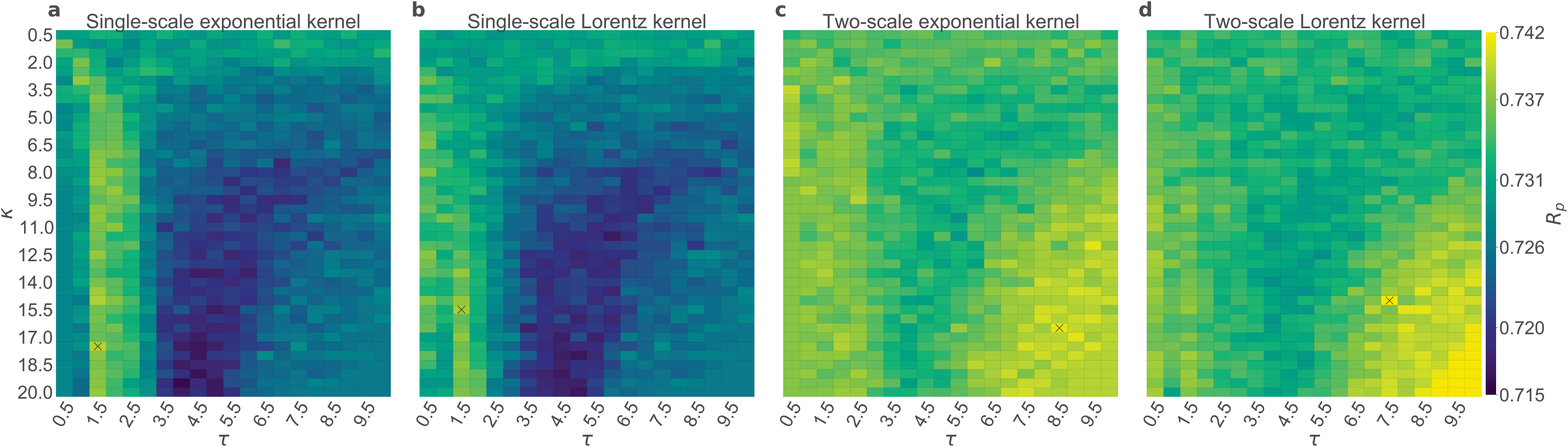}
    \caption{Optimized parameters for $^\text{ecif}\text{GGL}$ model for CASF--2007 benchmark. The best parameters locations are marked by ``x''. The optimal parameters for (a) single-scale exponential kernel model are $(\kappa,\tau)=(17.5,1.5)$ with the corresponding median $R_p=0.739$ and (b) single-scale Lorentz kernel model are $(\kappa,\tau)=(15.5,1.5)$ with corresponding median $R_p=0.738$. The optimal second kernel parameters for (c) two-scale exponential kernel model are $(\kappa,\tau)=(16.5,8.5)$ with the corresponding median $R_p=0.741$ and (d) two-scale Lorentz kernel model are $(\kappa,\tau)=(15.0,7.5)$ with the corresponding median $R_p=0.742$.}
    \label{fig:v2007_ecif_params}
\end{figure}

The hyperparameter optimization for the ECIF atom-type models is carried out in a similar fashion. Figure \ref{fig:v2007_ecif_params} displays the best parameters and the CV performance. We found that the best single-scale exponential model is $^\text{ecif}\text{GGL}_{17.5,1.5}^{\mathrm{E}}$ with median $R_p$=0.739 (Figure \ref{fig:v2007_ecif_params}a) and the best single-scale Lorentz kernel model is $^\text{ecif}\text{GGL}_{15.5,1.5}^{\mathrm{L}}$ with median $R_p$=0.738 (Figure \ref{fig:v2007_ecif_params}b). The best two-scale exponential kernel model is found to be $^\text{ecif}\text{GGL}_{17.5,1.5;16.5,8.5}^{\mathrm{E},\mathrm{E}}$ with median $R_p$=0.741 (Figure \ref{fig:v2007_ecif_params}c). Finally, (Figure \ref{fig:v2007_ecif_params}d), shows that the best two-scale Lorentz kernel model is $^\text{ecif}\text{GGL}_{15.5,1.5;15.0,7.5}^{\mathrm{L},\mathrm{L}}$ with median $R_p$=0.742.
  
Having optimized the models' hyperparameters, we now predict the binding affinity of the 195 complexes in the CASF-2007 test set. Just like in the previous two benchmarks, we first train each model using the smaller training set $S_R$ and produce a prediction for the test set from the average of 50 runs. Secondly, we use the bigger training set $S_G$ of this benchmark, which is the PDBbind v2007 general set excluding the CASF-2007 core set, to train the model and use the trained model to predict the binding affinity of the test set.

\begin{figure}[htbp]
\begin{center}
\includegraphics[width=0.95\textwidth]{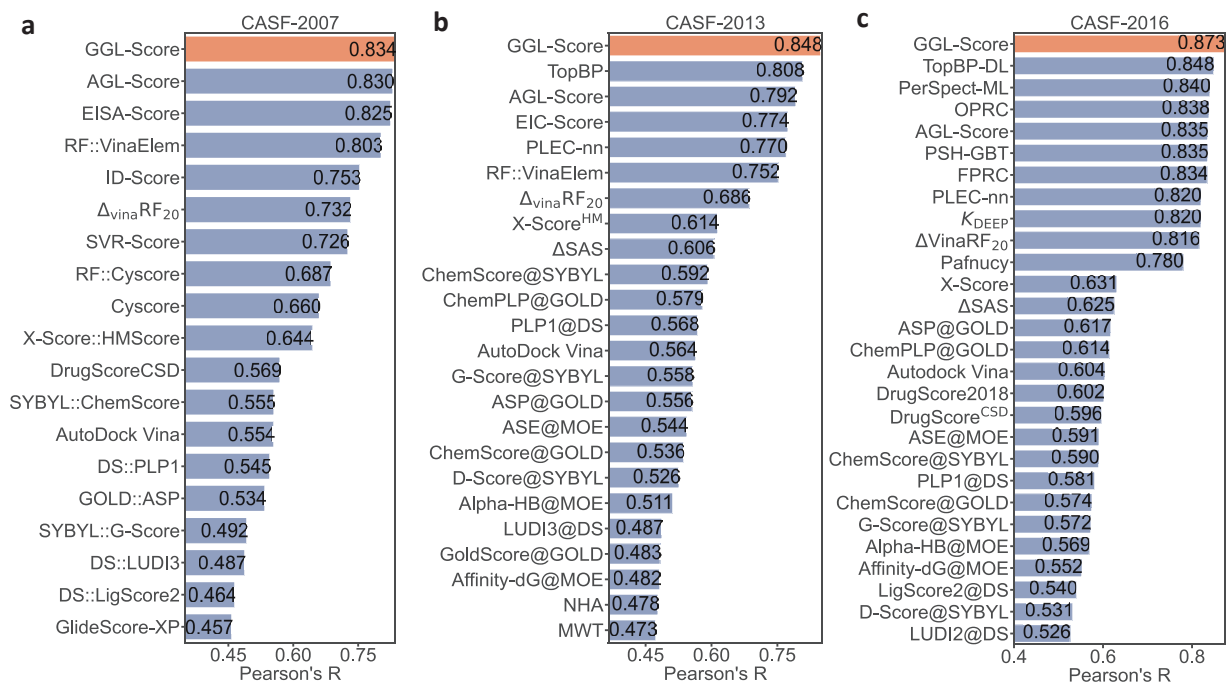}
\caption{Performance comparison of different scoring functions on CASF benchmarks. Our proposed model in this work, GGL-Score, is highlighted in red, and the rest is in purple. a) CASF--2007:  the performances of other methods taken from previous studies \cite{cheng2009comparative,ballester2010machine,li2013id,li2015improving,li2014substituting,cao2014improved,wang2017improving}. Our $^\text{sybyl}\text{GGL-Score}$ achieves $R_p$=0.834 b) CASF--2013: the other results are extracted from \cite{wang2017improving,li2014comparative,li2015improving}. Our $^\text{sybyl}\text{GGL-Score}$ achieves $R_p$=0.848. c) CASF--2016: our $^\text{sybyl}\text{GGL-Score}$ achieves $R_p$=0.873, other scoring functions are discussed in \cite{su2018comparative,stepniewska2018development,wang2017improving}.}
\label{fig:CASF_scoring_power}
\end{center}
\end{figure}

The performance of all our selected models for this benchmark is reported in Table \ref{tab:v2007_results}. We observe that all of these models perform significantly better when trained with the bigger training set $S_G$. Following a similar trend as in the previous two benchmarks, the SYBYL atom-type models of this benchmark consistently perform better than their ECIF atom-type counterparts. Also, the two-scale kernel model improves the performance compared to their single-scale versions. The best-performing model for this benchmark is the two-scale Lorentz kernel SYBYL atom-type model $^\text{sybyl}\text{GGL}_{13.5,0.5;13.0,9.5}^{\mathrm{L},\mathrm{L}}$ with Pearson\textquotesingle s correlation coefficient $R_p$=0.834. Moreover, Figure \ref{fig:CASF_scoring_power}a reveals the scoring power of our model in this benchmark. Our proposed GGL model stands at the top with reported $R_p$=0.834 while AGL-Score is the runner-up with $R_p$=0.830.

\begin{table}[htbp]
\begin{center}
\caption{Performance of various GGL models on CASF--2007 test set. \Bstrut}
\begin{tabular}{lcclc}
 \multicolumn{3}{@{}c@{}}{Pearson\textquotesingle s $R_p$ of single-scale Model} &
 \multicolumn{2}{@{}c@{}}{Pearson\textquotesingle s $R_p$ of two-scale Model}\Tstrut\Bstrut\\
\cline{1-3}\cline{4-5}
   Model & Trained with $S_R$ & Trained with $S_G$ & Model & Trained with $S_G$ \Tstrut\Bstrut\\
	\hline
    $^\text{sybyl}\text{GGL}_{2.5,0.5}^{\mathrm{E}}$ &0.803  &  0.824 & $^\text{sybyl}\text{GGL}_{2.5,0.5;19.0,9.0}^{\mathrm{E},\mathrm{E}}$ &  0.833\Tstrut\Bstrut\\
    $^\text{sybyl}\text{GGL}_{13.5,0.5}^{\mathrm{L}}$ &0.807   &  0.827 & $^\text{sybyl}\text{GGL}_{13.5,0.5;13.0,9.5}^{\mathrm{L},\mathrm{L}}$ & 0.834\Tstrut\Bstrut\\  
    \hline
     $^\text{ecif}\text{GGL}_{17.5,1.5}^{\mathrm{E}}$ &0.794  &   0.807 & $^\text{ecif}\text{GGL}_{17.5,1.5;16.5,8.5}^{\mathrm{E},\mathrm{E}}$ &  0.811\Tstrut\Bstrut\\
    $^\text{ecif}\text{GGL}_{15.5,1.5}^{\mathrm{L}}$ & 0.792  &  0.805 & $^\text{ecif}\text{GGL}_{15.5,1.5;15.0,7.5}^{\mathrm{L},\mathrm{L}}$ & 0.809\Tstrut\Bstrut\\
    \hline
\end{tabular}
\label{tab:v2007_results}
\end{center}
\end{table}


\section{Conclusion}
The binding affinity between a ligand and its receptor protein is a key component in structure-based drug design. Although significant progress has been made over the past decades, an accurate prediction of protein-ligand binding affinity remains a challenging task. Geometric graph theories are widely used in the study of molecular and biomolecular systems. Furthermore, the element-type graph coloring-based multiscale weighted colored graph (MWCG) approaches have particularly shown success in the task of binding affinity prediction \cite{nguyen2017rigidity, nguyen2019agl}. On the other hand, SYBYL atom-type interaction and extended connectivity interactive features (ECIF) have enjoyed their success in molecular property prediction \cite{ballester2014does,sanchez2021extended}. Therefore, with an aim to develop robust and reliable scoring functions for large and diverse protein-ligand datasets, the present work combines the graph learning model and extended atom types to give rise to novel geometric graph theory-based multiscale weighted colored graph (MWCG) descriptors for the protein-ligand complex where the graph coloring is based on SYBYL atom-type and ECIF atom-type interactions. By pairing with the gradient boosting decision tree (GBDT) machine learning algorithm, our approach results in two different methods, namely $^\text{sybyl}\text{GGL}$-Score and $^\text{ecif}\text{GGL}$-Score. We explore the optimal hyperparameters of our models using a five-fold cross-validation on the training set of three commonly used benchmarks in drug design area, namely CASF-2007 \cite{cheng2009comparative}, CASF-2013 \cite{li2014comparative}, and CASF-2016 \cite{su2018comparative}. For the binding affinity prediction task of each benchmark's test set complexes, we consider two training sets-- the refined set excluding the core set and the general set excluding the core set. Our model performs significantly better in each benchmark when trained with the larger training set. It is also found that the SYBYL atom-type models $^\text{sybyl}\text{GGL}$-Score outperform the ECIF atom-type models $^\text{ecif}\text{GGL}$-Score in most cases.

To demonstrate the scoring power of the proposed models, many state-of-the-art scoring functions are considered in each benchmark.
Impressively, our  $^\text{sybyl}\text{GGL}$-Score outperforms other models by a wide margin in all three PDBbind benchmarks. In addition to the accuracy and robustness, our model is computationally inexpensive-- the only required structural input is the atom types and coordinates. Moreover, our model can be applied in a vast majority of molecular property predictions such as toxicity, solubility, protein mutation, protein folding, and protein-nucleic acid interactions.

\setcounter{table}{0}
\renewcommand{\thetable}{A\arabic{table}}
\setcounter{figure}{0}
\renewcommand\thefigure{A\arabic{figure}}

\section{Appendix A}\label{appendix}
\begin{figure}[htbp]
    \centering
    \includegraphics[width=1\linewidth]{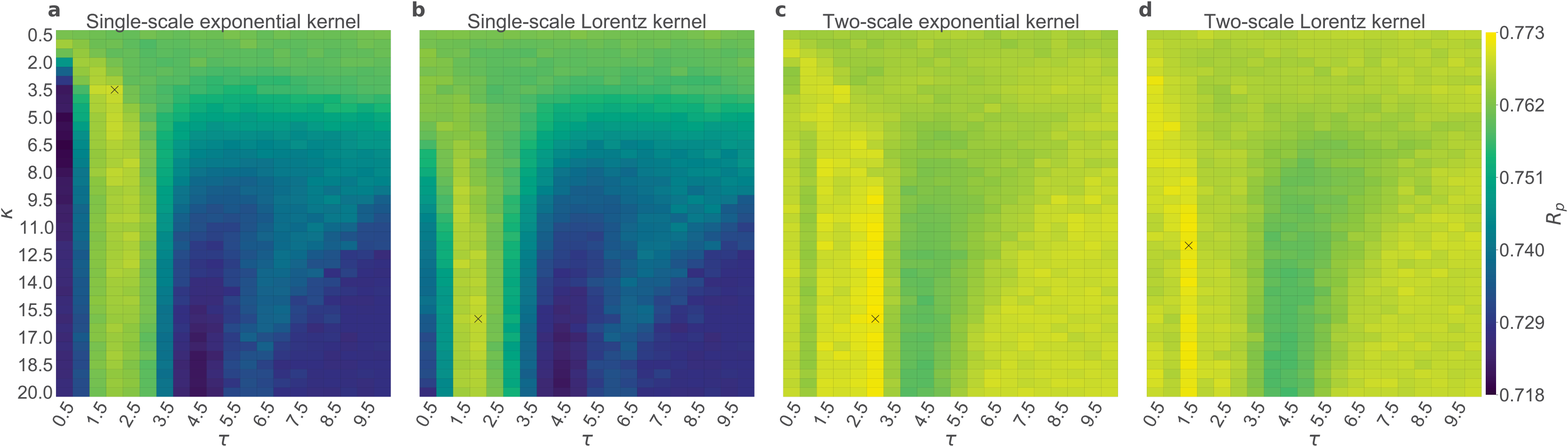}
    \caption{Optimized parameters for basic $\text{GGL}$ model for CASF--2016 benchmark. The best parameters locations are marked by ``x''. The optimal parameters for (a) single-scale exponential kernel model are $(\kappa,\tau)=(3.5,2.0)$ with the corresponding median $R_p=0.769$ and (b) single-scale Lorentz kernel model are $(\kappa,\tau)=(16.0,2.0)$ with corresponding median $R_p=0.769$. The optimal second kernel parameters for (c) two-scale exponential kernel model are $(\kappa,\tau)=(16.0,3.0)$ with the corresponding median $R_p=0.773$ and (d) two-scale Lorentz kernel model are $(\kappa,\tau)=(12.0,1.5)$ with the corresponding median $R_p=0.773$.}
    \label{fig:v2016_element_type_kernel_params}
\end{figure}

In this section, we explore the performance of the basic geometric graph approach model that considers element-type interactions presented in \cite{nguyen2017rigidity} using the bigger training set $S_G$, i.e. the general set excluding the core set of each benchmark. We carry out a similar experiment as we did for our present model. For simplicity, we use the notation $\text{GGL}_{\kappa,\tau}^{\alpha}$ for a single-scale kernel and $\text{GGL}_{\kappa_1,\tau_1;\kappa_2,\tau_2}^{\alpha_1, \alpha_2}$ for a two-scale kernel basic element-type geometric graph learning model. To find the optimized parameters for each benchmark, we carry out a five-fold CV on the training set $S_R$ i.e. the refined set excluding the core set of the corresponding benchmark. For CASF--2016 benchmark, the best single kernel models are found to be $\text{GGL}_{3.5,2.0}^{\mathrm{E}}$ (Figure \ref{fig:v2016_element_type_kernel_params}a)and $\text{GGL}_{16.0,2.0}^{\mathrm{L}}$ (Figure \ref{fig:v2016_element_type_kernel_params}b) with median Pearson correlation $R_p$=0.769 for both models. The best two kernel models for CASF--2016 are  $\text{GGL}_{3.5,2.0;16.0,3.0}^{\mathrm{E},\mathrm{E}}$ (Figure \ref{fig:v2016_element_type_kernel_params}c) and $\text{GGL}_{16.0,2.0;12.0,1.5}^{\mathrm{L},\mathrm{L}}$ (Figure \ref{fig:v2016_element_type_kernel_params}d) with median $R_p$=0.773 for both models. The performance of all models on the test set of CASF--2016 benchmark are reported in Table \ref{tab:standard_atom_type_results}. It is interesting to find that the performance of each model improved significantly when trained with the bigger training data $S_G$ i.e. PDBbind v2016 general set excluding the core set. The best performing model for this benchmark is the two-scale exponential kernel model $\text{GGL}_{3.5,2.0;16.0,3.0}^{\mathrm{E},\mathrm{E}}$ with $R_p$=0.859. We note that both of our proposed GGL models, SYBYL atom-type and ECIF atom-type model, perform promisingly better (with reported $R_p$=0.873 and 0.868 respectively) than the basic element-type model.  

\begin{figure}[htbp]
    \centering
    \includegraphics[width=1\linewidth]{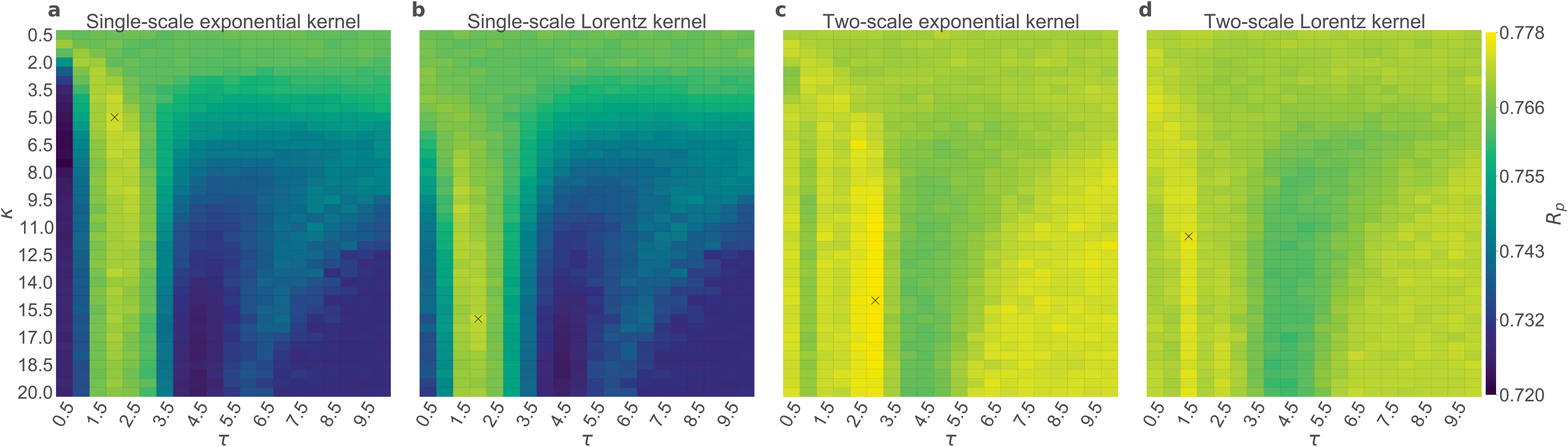}
    \caption{Optimized parameters for basic $\text{GGL}$ model for CASF--2013 benchmark. The best parameters locations are marked by ``x''. The optimal parameters for (a) single-scale exponential kernel model are $(\kappa,\tau)=(5.0,2.0)$ with the corresponding median $R_p=0.774$ and (b) single-scale Lorentz kernel model are $(\kappa,\tau)=(16.0,2.0)$ with corresponding median $R_p=0.773$. The optimal second kernel parameters for (c) two-scale exponential kernel model are $(\kappa,\tau)=(15.0,3.0)$ with the corresponding median $R_p=0.778$ and (d) two-scale Lorentz kernel model are $(\kappa,\tau)=(11.5,1.5)$ with the corresponding median $R_p=0.776$.}
    \label{fig:v2013_element_type_kernel_params}
\end{figure}

The CV performance of CASF--2013 benchmark (Figure \ref{fig:v2013_element_type_kernel_params}), reveals that the best models for this benchmark are $\text{GGL}_{5.0,2.0}^{\mathrm{E}}$ and    $\text{GGL}_{16.0,2.0}^{\mathrm{L}}$ for the single kernel, and $\text{GGL}_{5.0,2.0;15.0,3.0}^{\mathrm{E},\mathrm{E}}$ and $\text{GGL}_{16.0,2.0;11.5,1.5}^{\mathrm{L},\mathrm{L}}$ for two kernels, with median $R_p$= 0.774, 0.773, 0.778, and 0.776, respectively. Table \ref{tab:standard_atom_type_results} indicates that the use of the bigger training set $S_G$ improved the performance of these models. The best performing model for this benchmark is the single-scale exponential model $\text{GGL}_{5.0,2.0}^{\mathrm{E}}$ with reported $R_p$=0.821. However, our proposed SYBYL atom-type GGL model outperforms the basic GGL model by a huge margin with reported $R_p$=0.848 for this benchmark.

\begin{figure}[htbp]
    \centering
    \includegraphics[width=1\linewidth]{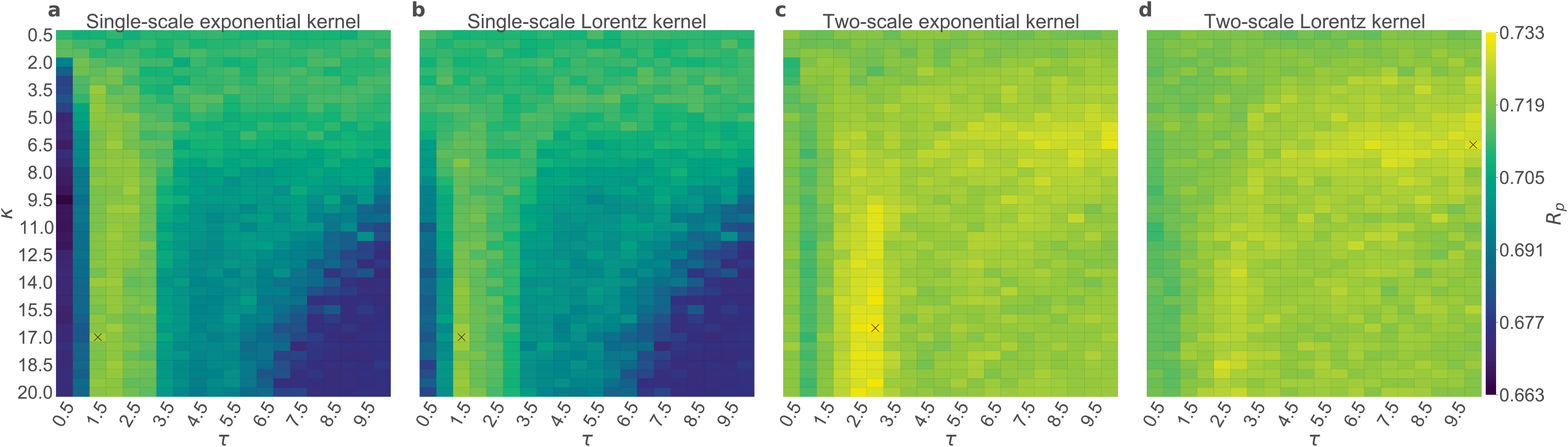}
    \caption{Optimized parameters for basic $\text{GGL}$ model for CASF--2007 benchmark. The best parameters locations are marked by ``x''. The optimal parameters for (a) single-scale exponential kernel model are $(\kappa,\tau)=(17.0,1.5)$ with the corresponding median $R_p=0.724$ and (b) single-scale Lorentz kernel model are $(\kappa,\tau)=(17.0,1.5)$ with corresponding median $R_p=0.724$. The optimal second kernel parameters for (c) two-scale exponential kernel model are $(\kappa,\tau)=(16.5,3.0)$ with the corresponding median $R_p=0.733$ and (d) two-scale Lorentz kernel model are $(\kappa,\tau)=(6.5,10.0)$ with the corresponding median $R_p=0.730$.}
    \label{fig:v2007_element_type_kernel_params}
\end{figure}

Figure \ref{fig:v2007_element_type_kernel_params} plots the CV performance for CASF--2007 benchmark. The best models for this benchmark are $\text{GGL}_{17.0,1.5}^{\mathrm{E}}$ and    $\text{GGL}_{17.0,1.5}^{\mathrm{L}}$ for the single kernel, and $\text{GGL}_{17.0,1.5;16.5,3.0}^{\mathrm{E},\mathrm{E}}$ and $\text{GGL}_{17.0,1.5;6.5,10.0}^{\mathrm{L},\mathrm{L}}$ for two kernels, with median $R_p$= 0.724, 0.724, 0.733, and 0.730, respectively. The performance of these models is presented in Table \ref{tab:standard_atom_type_results}. We observe that the use of the bigger training data significantly improves the performance of each model for this benchmark as well. While the best performing basic GGL model for this benchmark is the two-scale exponential kernel model $\text{GGL}_{17.0,1.5;16.5,3.0}^{\mathrm{E},\mathrm{E}}$ with Pearson's $R_p$=0.833, our proposed SYBYL atom-type GGL model for this benchmark perform slightly better with $R_p$=0.834.



\begin{table}[htbp]
\begin{center}
\caption{Performance of various basic GGL models on all benchmark test sets. \Bstrut}
\begin{tabular}{l|lcclc}
 \multicolumn{1}{c}{} & \multicolumn{3}{@{}c@{}}{Pearson\textquotesingle s $R_p$ of single-scale Model} &
 \multicolumn{2}{@{}c@{}}{Pearson\textquotesingle s $R_p$ of two-scale Model}\Tstrut\Bstrut\\
\cline{2-4}\cline{5-6}
\multicolumn{1}{c}{} & Model & Trained with $S_R$ & Trained with $S_G$ & Model & Trained with $S_G$ \Tstrut\Bstrut\\
	\hline
 \multirow{2}{*}{CASF--2016}
    & $\text{GGL}_{3.5,2.0}^{\mathrm{E}}$  & 0.843 & 0.856 & $\text{GGL}_{3.5,2.0;16.0,3.0}^{\mathrm{E},\mathrm{E}}$ &  0.859\Tstrut\Bstrut\\
   & $\text{GGL}_{16.0,2.0}^{\mathrm{L}}$   & 0.839 & 0.848 & $\text{GGL}_{16.0,2.0;12.0,1.5}^{\mathrm{L},\mathrm{L}}$ & 0.856\Tstrut\Bstrut\\  
    \hline
 \multirow{2}{*}{CASF--2013}
    & $\text{GGL}_{5.0,2.0}^{\mathrm{E}}$  & 0.794 & 0.821 & $\text{GGL}_{5.0,2.0;15.0,3.0}^{\mathrm{E},\mathrm{E}}$ &  0.818\Tstrut\Bstrut\\
   & $\text{GGL}_{16.0,2.0}^{\mathrm{L}}$   & 0.793 & 0.809 & $\text{GGL}_{16.0,2.0;11.5,1.5}^{\mathrm{L},\mathrm{L}}$ & 0.818\Tstrut\Bstrut\\  
    \hline
 \multirow{2}{*}{CASF--2007}
    & $\text{GGL}_{17.0,1.5}^{\mathrm{E}}$  & 0.809 & 0.828 & $\text{GGL}_{17.0,1.5;16.5,3.0}^{\mathrm{E},\mathrm{E}}$ &  0.833\Tstrut\Bstrut\\
   & $\text{GGL}_{17.0,1.5}^{\mathrm{L}}$   & 0.815 & 0.830 & $\text{GGL}_{17.0,1.5;6.5,10.0}^{\mathrm{L},\mathrm{L}}$ & 0.830\Tstrut\Bstrut\\  
    \hline
\end{tabular}
\label{tab:standard_atom_type_results}
\end{center}
\end{table}

\section{Data and Software Availability}
The source code is available at Github: \url{https://github.com/NguyenLabUKY/GGL-ETA-Score}.

 \section{Competing interests}
 No competing interest is declared.

\section{Acknowledgments}
This work is supported in part by funds from the National Science Foundation (NSF: \# 2053284 and \# 2151802), and University of Kentucky Startup Fund.


\end{document}